\documentstyle[12pt]{article}
\begin{document}
\pagestyle{empty}
\hspace*{12.4cm}IU-MSTP/37\\
\hspace*{13cm}hep-lat/9910030 \\
\hspace*{13cm}September, 1999
\begin{center}
 {\Large\bf Application of Noncommutative \\
Differential Geometry on Lattice to Anomaly\\  
\vskip .3cm
Analysis 
in Abelian Lattice Gauge Theory}
\end{center}

\vspace*{1cm}
\def\thefootnote{\fnsymbol{footnote}}
\begin{center}{\sc Takanori Fujiwara,} 
{\sc Hiroshi Suzuki}
and {\sc Ke Wu}\footnote{On leave of absence from 
Institute of Theoretical Physics, Academia Sinica, P.O.Box 2735, Beijing 
100080, China}
\end{center}
\vspace*{0.2cm}
\begin{center}
{\it Department of Mathematical Sciences, Ibaraki University,
Mito 310-8512, Japan}
\end{center}
\vfill
\begin{center}
{\large\sc Abstract}
\end{center}
\noindent
The chiral anomaly in lattice abelian gauge theory 
 is investigated by applying the geometric and topological 
method in noncommutative differential geometry(NCDG).  A new kind of double complex and descent 
equation are proposed on infinite hypercubic lattice in arbitrary even dimensional Euclidean space, in 
the framework of NCDG. Using the general solutions to proposed descent equation, we derive the chiral anomaly
in Abelian lattice gauge theory. The topological origin of anomaly is 
nothing but the Chern 
classes in NCDG.



\newpage
\pagestyle{plain}

\section{Introduction}

\label{sec:intro}
\setcounter{equation}{0}

\noindent
There is an important no-go theorem in lattice gauge theory proved by
Nielson and Ninomiya (NN)\cite{NN}  which states
that it is in fact impossible to formulate a theory
of chiral fermion on lattice with a few plausible assumptions. 
The construction of chiral field theories on lattice thus appears to be more difficult.
Since then lots of efforts have been done in order to escape NN theorem, one of them was suggested  by 
Ginsparg and Wilson \cite{GW} many years ago. They modified 
the chirality condition of Dirac operator $D$, one of the assumptions 
in NN theorem, to
  the so called Ginsparg-Wilson  (GW) relation:
\begin{eqnarray}
  \label{gw}
  \gamma_5 D + D \gamma_5 = a D\gamma_5 D,
\end{eqnarray}
where $a$ is the lattice spacing, taken to be unity in the present paper.

Recently, there have been remarkable developments regarding this
problem.\cite{Neu, Hass, HLN, Lus1}  It has been shown that the chiral symmetry may be preserved in 
lattice gauge theory, 
at least to some extent, 
if the lattice Dirac operator takes a particular form. Although the Dirac operators proposed by them are 
very complicated, but all of them satisfy the GW relation, gauge invariance and locality. 
It was soon recognized that GW relation
implies the index theorem \cite{Hass,HLN, Lus1} and the exact chiral
symmetry\cite{Lus1} of the fermion action. The
chiral or axial anomaly arises from the non-invariance of the 
fermion integration measure under the chiral transformations. In terms of the GW Dirac operator $D$  it 
 can be written as
\begin{eqnarray}
        \label{anom}
   q(x)=- \frac {1}{2} {\rm tr}{\gamma_5 D(x,x)}.
\end{eqnarray}

From the property of Dirac operator $D$ satisfying GW relation, the anomaly 
is gauge invariant and local (in the sense that it is locally dependent on gauge field). 
Moreover its variation under any 
local deformation of gauge field satisfies
\begin{eqnarray}
        \label{topo}
   \sum_x \delta q(x)&=&0 ~.
\end{eqnarray}
which reflects the topological nature of the anomaly. 

The understanding of structure of the gauge anomaly on the lattice with finite spacing is quite important. 
There are many papers \cite{many} appeared recently to  study the properties about 
chiral anomaly on lattice,  such as its perturbative property, continuum limit and etc. 
Among them  L\"uscher derived a theorem \cite{Lus2}
to show the explicit form of  the chiral anomaly $q(x)$ 
in abelian gauge theory on four-dimensional euclidean infinite hypercubic lattice,
\begin{eqnarray}
        \label{anom1}
        q(x)=\alpha +\beta_{\mu\nu} F_{\mu\nu}(x)
   +\gamma\varepsilon_{\mu\nu\rho\sigma}
   F_{\mu\nu}(x)F_{\rho\sigma}(x+\widehat\mu+\widehat\nu)
   +\partial_\mu^*k_\mu(x),
\end{eqnarray}
where some notation will be explained in next section.
Furthermore L\"uscher proved that the 
gauge invariant lattice formulation of Abelian chiral 
gauge theory satisfied all other physical requirements could be exist.\cite{Lus3}

L\"uscher's theorem  has been extended to arbitrary even dimensional regular lattice by us 
using noncommutative  geometry. In one of our papers \cite{FSW} 
BRST analysis, which was used to get the expression of chiral anomaly as in continuum theory, 
is extended  to abelian lattice gauge theory, 
once the exterior derivative on lattice  
with similar nilpotency and Leibniz rule as usual differential geometry is obtained. 
A double complex and descent equation, as well as its general solution are used to derive 
the anomaly in another paper \cite{FSW1}. 
In this paper we will continue our research on chiral anomaly of lattice gauge theory. In terms of 
a new kind of double complex and its descent equation in 
 non-commutative differential geometry (NCDG) the chiral anomaly will be derived. 
The paper is organized as follows. Firstly,
some basic result of NCDG  on lattice will be reviewed in Section 2. Then a new kind of double complex and 
descent equation are proposed, and the general solution to the descent equation will be studied in Section 3.
 The chiral anomalies are derived in Section 4. Finally conclusion and remarks are included in Section 5.

\section{Noncommutative differential geometry \\
on D-dimensional lattice ${\rm Z}^D$}
\label{sec:NCDC}
\setcounter{equation}{0}

NCDG is of long history and wide applications in physics, especially in quantum physics. In fact,
one can trace back to Heisenberg and Dirac at the early stage of this 
century, when they founded the 
quantum mechanics. They deformed the commutative 
classical phase space to  quantum phase space which is 
noncommutative. In quantum mechanics, the physical quantities have to be quantized to be operators
in Hilbert space. It is such representation that motivated von Neumann to study the operator algebras 
which in turn inspired the foundation of noncommutative geometry. The main method in 
NCDG by Connes \cite{Conn} is to study the algebraic object like function
algebra ~$\cal A(\rm M)$~ defined on  manifold ~$\rm M$~ rather than the geometric object 
such as manifold M itself. So that it could be used to 
discuss some geometric object which 
can not be treated in the usual way, such as the discrete set, or quantum group. 
The NCDG on lattice used in this paper is the simple case of noncommutative differential 
calculus on discrete set.
Recently its application on  lattice gauge theory has got some progress\cite {Lus2}, although most of 
the involved mathematics have appeared 
before\cite {Conn, Sit}. Some basic results are reviewed in this section without proof.

The $D$-dimensional infinite hypercubic regular lattice  ~${\rm Z^D}$ could be 
understood as the discretization of continuous Euclidean space ~$\rm M^D$. 
The discreteness can not be expressed 
properly
by usual differential geometry. Instead it can be well described by noncommutative geometry.
In this formalism, first remarkable property is that there are two types of vector field defined on the 
algebra $\cal A({\rm Z}^{\rm D})$, $\cal A({\rm Z}^{\rm D})$ being the algebra of function defined on ~${\rm Z}^D$: the forward difference 
operator ~$\partial_\mu$~ and the backward difference
operator~$\partial_\mu^*$, playing the role of  the ordinary derivative $\partial_x$ on $\cal A(\rm M)$,
\begin{eqnarray}
  \label{forw}
   \partial_\mu f(x)=f(x+\widehat\mu)-f(x),\nonumber\\
   \partial_\mu^* f(x)=f(x)-f(x-\widehat\mu),
\end{eqnarray}
where ~$f(x)$~ is an element in ~$\cal A({\rm Z}^{\rm D})$~ i.e. arbitrary function on the lattice and
~$\widehat\mu$~ stands for the unit vector in direction~$\mu$.    

 In order to get  the full differential calculus on ~$\cal A({\rm Z^D})$~ 
one needs to define differential forms on it. The basis of 1-form on it 
are abstract objects ~$dx_1,dx_2,\cdots,dx_D,$ satisfing the
Grassmann algebra relation:
\begin{eqnarray}
  \label{diff}
    dx_\mu dx_\nu=-dx_\nu dx_\mu.
\end{eqnarray}
Then we can define exterior derivative operator ~$d$ on ~$\cal A({\rm Z^D})$ with forward difference operator as
\begin{eqnarray}
  \label{diff1}
df(x)=\partial_\mu f(x)dx_\mu,
\end{eqnarray}
where the summation on repeated indices is understood. 
For generic ~$n$-form,
\begin{eqnarray}
  \label{diff2}
   f(x)={1\over n!}\,f_{\mu_1\cdots\mu_n}(x)
   \,dx_{\mu_1}\cdots dx_{\mu_n},
\end{eqnarray}
where the coefficient ~$ f_{\mu_1\cdots\mu_n}(x)$ is antisymmetric tensor with rank ~$n$, 
the exterior derivative operator ~$d$ maps it to ~$(n+1)$-form as
\begin{eqnarray}
  \label{diff3}
   df(x)={1\over n!}\,\partial_\mu f_{\mu_1\cdots\mu_n}(x)
   \,dx_\mu dx_{\mu_1}\cdots dx_{\mu_n}.
\end{eqnarray}
It is easy to notice that the difference operators 
do not satisfy the ordinary Leibniz rule as usual derivative operator. 
 This is an important difference between 
commutative and noncommutative geometry. In order to get correct differential calculus 
on ~$\cal A({\rm Z^D})$, one should remedy this shortage of Leibniz rule for difference operator. 
One try to define the exterior derivative operator $d$, instead of difference operator,   to satisfy 
Leibniz rule. 
There is an important lemma about exterior derivative operator ~$d$ on ~$\cal A({\rm Z^D})$ \cite{Sit} 
in order to get the correct Leibniz rule for ~$d$.
\vskip .5cm
\noindent{\bf Lemma 1:} The exterior derivative operator $d$ on $\cal A({\rm Z^D})$ satisfy the nilpotency
$d^2=0$ and Leibniz rule 
\begin{eqnarray}
  \label{leib}
d[f(x)g(x)]=df(x)g(x)+(-1)^nf(x)dg(x),
\end{eqnarray}
where $f(x)$ is $n$-form, if and only if the forms and functions 
satisfy the  following conditions:
\begin{eqnarray}
  \label{ncomm}
dx_\mu dx_\nu=-dx_\nu dx_\mu, \nonumber\\
dx_\mu f(x)=f(x+\widehat\mu)\,dx_\mu.
\end{eqnarray}
\vskip .5cm
The last relation shows that the function and 1-forms can not be commutative in order to have the right
Leibniz rule, s.t. the noncommutativity appears. This is the essence of NCDG on regular lattice 
and represents the discreteness of lattice.

We can also introduce  the gauge potential 1-form and
the field strength 2-form on the lattice in terms of NCDG once we get all the form space and exterior 
derivative $d$ acting on it satisfying the nilpotency and the Leibniz rule,
\begin{eqnarray}
  \label{gauge}
   A(x)=A_\mu(x)\,dx_\mu, ~~~~~~~
   F(x)=dA(x)   =\frac {1}{2} F_{\mu\nu}(x)\,dx_\mu dx_\nu
\end{eqnarray}
where $ F_{\mu\nu}(x)=\partial_\mu A_\nu(x)-\partial_\mu A_\nu(x)$ and $d$ is the exterior derivative operator
on $\cal A({\rm Z^D})$.
Note that the Bianchi identity~$dF(x)=0$ holds from the nilpotency of $d$.
Under gauge transformation the gauge potential 1-form $A(x)$ becomes $A(x)+d\lambda(x)$, then one can 
know the  field strength 2-form $F(x)$ is gauge invariant. 
Although both the gauge potential and the field strength have the same form as ones in ordinary 
gauge theory, they have different meaning as it is defined on lattice using NCDG and 
called the noncommutative gauge theory on lattice, 
where the noncommutavity are widely used.

In usual lattice gauge theory the fundamental gauge field is link variable 
~$U_\mu(x)$, which  is defined on the link connecting 
the points~$x$ and~$x+\widehat\mu$ and taking value on gauge group $G$. 
We restrict our discussion on abelian case ~$G=U(1)$ in this paper. 
Gauge transformation acting on gauge fields is through 
\begin{eqnarray}
  \label{tran}
U_\mu(x)   \to      \Lambda(x) U_\mu(x){\Lambda(x+\widehat\mu)}^{-1}.           
\end{eqnarray}
Where the  parameter $\Lambda(x)$ of gauge transformations also take the values on $U(1)$.
Unlike the noncommutative gauge theory on lattice the gauge field strength 
~$F_{\mu\nu}(x)$ on lattice gauge theory 
is given by the plaquette variable
\begin{eqnarray}
  \label{link}
   F_{\mu\nu}(x)={1\over i}\ln U_\mu(x)U_\nu(x+\widehat\mu)
   U_\mu(x+\widehat\nu)^{-1}U_\nu(x)^{-1}.
\end{eqnarray}
 We  assume the
``admissibility" condition for the gauge field configuration\cite{Lus2}, i.e.:
\begin{eqnarray}
  \label{admi}
   \sup_{x,\mu,\nu}|F_{\mu\nu}(x)|<\epsilon,
\end{eqnarray}
where $\epsilon$ is a fixed constant $0<\epsilon<\pi/3$. 
Then we will get the relation of noncommutative gauge theory on lattice and lattice gauge theory 
by the lemma from L\"uscher \cite{Lus2}.

\vskip .5cm
\noindent{\bf Lemma 2:} Suppose $U_\mu(x)$ is admissible $U(1)$ gauge field. then there exists a 
vector field $A_\mu(x)$ such that:
\begin{eqnarray}
  \label{admi1}
U_\mu(x)=e^{iA_\mu(x)},~~~~F_{\mu\nu}(x)=\partial_\mu A_\nu(x)-\partial_\mu A_\nu(x).
\end{eqnarray}
\vskip .5cm
\noindent This lemma shows that the noncommutative gauge theory defined by NCDG on lattice is equivalent to lattice 
gauge theory at least in the abelian case. 

In noncommutative gauge theory on lattice ${\rm Z}^D$, the Chern classes of that gauge theory  
could be defined as:
\begin{eqnarray}
  \label{chern}
c_k=F^k=F F\cdots F,~~~~ k=1,2,\cdots,\frac {D}{2}.
\end{eqnarray}
Although it looks like the Chern class in usual differential geometry, in fact, it is a different one   
\begin{eqnarray}
  \label{chern1}
c_k&=&F_{\mu_1\nu_1}(x)
   F_{\mu_2\nu_2}(x+\widehat\mu_1+\widehat\nu_1)\cdots
   F_{\mu_k\nu_k}(x+\widehat\mu_1+\widehat\nu_1+\cdots
    +\widehat\mu_{k-1}+\widehat\nu_{k-1})\nonumber \\
&&dx_{\mu_1}dx_{\nu_1}dx_{\mu_2}dx_{\nu_2}\cdots dx_{\mu_k}dx_{\nu_k}
\end{eqnarray}
One can easily prove that $c_k$ is gauge invariant and closed $dc_k=0$.

In order to understand the topological properties of 
D-dimensional lattice ${\rm Z}^D$ i.e. the cohomology properties of $\cal A({\rm Z^D})$ 
in terms of NCDG, one needs the Poincar\'e lemma.
For the simplicity  we will restrict our discussion to the $k$-forms with 
compact support or exponentially decaying coefficients (this is corresponding 
to the  locality in lattice gauge theory) 
and denote the linear space of all $k$-forms 
as $\Omega_k$. The Poincar\'e lemma is as follows \cite {Lus2},
\vskip .5cm
\noindent{\bf Lemma 3 (Poincar\'e Lemma }for $d$): Let $f \in \Omega_k $ be a $k$-form satisfying:
$df(x)=0$ and $\sum_x f(x)=0$ for $k=D$. Then there exists a form $g(x)\in\Omega_{k-1}$ such that
$f(x)=d g(x)$.
\vskip .5cm
In the proof of this lemma, the locality plays a crucial role. If the  reference point  chosen 
in the proof is located in the compact support of $f(x)$, then  $g(x)$ is supported on the same rectangular 
block of lattice as $f(x)$.  
The construction of the form $g(x)$ is explicit and its coefficients are just some 
particular linear combinations of the coefficients of $f(x)$.

In form space $\Omega_k$ the inner product, corresponding to  
the metric in D-dimensional lattice ${\rm Z}^D$, 
can also be defined as:
\begin{eqnarray}
  \label{inner} 
<dx_\mu, dx_\nu>&=&\delta_{\mu,\nu},\nonumber \\
<dx_{\mu_1}\cdots dx_{\mu_k}, dx_{\nu_1}\cdots dx_{\nu_l}>&=&\delta_{k,l} 
~{\rm det}(<dx_{\mu_i}, dx_{\nu_j}>_{ij}).
\end{eqnarray}
Then the divergence operator $d^*:\,\Omega_k\to\Omega_{k-1}$ defined by the backward difference operator 
\begin{eqnarray}
  \label{diver}
   d^*f(x)={1\over (k-1)!}\,\partial^*_\mu f_{\mu\mu_2\cdots\mu_k}(x)
   \,dx_{\mu_2}\cdots dx_{\mu_n},
\end{eqnarray}
is nothing but the dual operator of ~$-d$~ in the Hilbert space 
$\Omega_{\bullet} = \sum_{k=0}^{k=D}\Omega_k$ with inner product  ~~$\sum_x<\cdot~,~\cdot>$. 
It also is nilpotent
$(d^*)^2=0$ and hence there will be another Poincar\'e lemma for $d^*$\cite{Lus2}.

\vskip .5cm
\noindent{\bf Lemma 4 (Poincar\'e Lemma }for $d^*$) Let $f \in \Omega_k $ be a $k$-form satisfying:
$d^* f(x)=0$ and $\sum_x f(x)=0$ for $k=0$. Then there exists a form $g(x)\in\Omega_{k+1}$ such that
$f(x)=d^* g(x)$.
\vskip .5cm
\noindent The proof of this lemma  is almost the same as that of the {\bf Lemma 3}. 
Also the locality plays an important role. 
 
One should notice that if one naively applies these lemmas 
to lattice gauge theory with a particular choice of the reference point, then the translation 
invariance will be lost as well as the locality. In order to overcome this difficulty, the ``bi-local" field 
$f(x,y)$ was used in L\"uscher's paper \cite {Lus2}, when one applies the Poincar\'e lemma to $x$ then $y$ is the
reference point and vice versa. The ``bi-local" means that the composite field $f(x,y)$ decreases 
at least exponentially as  ~$|x-y|\to\infty$. In this way, the locality is resumed, and the Poincar\'e 
Lemma can be applied to lattice gauge theory without problem. 

Therefore we should work on the ``bi-local" fields. Let us introduce the non-commutative differential 
calculus on the  algebra of all ``bi-local" field, which is the functions of both $x$ and $y$ defined on 
$\rm Z^D \times \rm Z^D$. Besides two types of vectors of forward difference operator 
 ~$\partial_\mu$ and backward difference
operator~$\partial_\mu^*$ for $x$, there is one more set of forward and backward 
difference operators acting on $y$, defined as,
\begin{eqnarray}
  \label{diffy}
    \partial^y_\mu f(x,y)=f(x,y)\overleftarrow\partial_\mu=f(x,y+\hat\mu)-f(x,y)~, \nonumber\\
  \partial^{*y }_\mu f(x,y)=  f(x,y)\overleftarrow\partial{}^*_\mu=f(x,y)-f(x,y-\hat\mu)~.
\end{eqnarray}
Furthermore  we need another copy of the basis of exterior 1-forms $dy_\mu (\mu=1,2,\cdots,D)$ satisfying 
\begin{eqnarray}
\label{diffy1}
dy_\mu dy_\nu= -dy_\nu dy_\mu,~~~~ 
dy_\mu f(x,y)=f(x,y+\widehat\mu)\,dy_\mu.
\end{eqnarray}
In a similar way we will get the exterior derivative operator $d_y$, exterior divergence operator $d^*_y$ 
and all the noncommutative differential calculus with respect to $y$.

A differential $(k,l)$-form on ${\rm Z}^D\times{\rm Z}^D$ is defined by 
\begin{eqnarray}
  \label{dklform}
  f=\frac{1}{k!l!}f_{\mu_1\cdots\mu_k;\nu_1\cdots\nu_l}(x,y){\rm d}x_{\mu_1}\cdots
  {\rm d}x_{\mu_k}{\rm d}y_{\nu_1}\cdots {\rm d}y_{\nu_l}~,
\end{eqnarray}
where $f_{\mu_1\cdots\mu_k;\nu_1\cdots\nu_l}(x,y)$ is completely antisymmetric 
in $\mu_1,\cdots,\mu_k$ and in $\nu_1,\cdots,\nu_l$, separately, also it  is 
assumed to have compact support on ${\bf Z}^D\times\{y\}$ and $\{x\}\times{\bf Z}^D$ or
to decrease at least exponentially as  $|x-y|\to\infty$ in order to apply Poincar\'e lemma on it. 
The vector space of ``bi-local" field $(k,l)$-forms is denoted by $\Omega_{k,l}$. 

The exterior differential with respect to $x$ or $y$ is denoted by $d_x$ or $d_y$, acting 
on the $(k,l)$-form (\ref{dklform}) is, 
\begin{eqnarray}
  \label{extdxdy}
  && {d}_xf=\frac{1}{k!l!}\partial_\mu f_{\mu_1\cdots\mu_k;\nu_1\cdots\nu_l}(x,y)
  { d}x_{\mu}{ d}x_{\mu_1}\cdots
  {d}x_{\mu_k}{ d}y_{\nu_1}\cdots { d}y_{\nu_l}~, \nonumber\\
  && { d}_yf=\frac{(-1)^k}{k!l!}f_{\mu_1\cdots\mu_k;\nu_1\cdots\nu_l}(x,y)
  \overleftarrow\partial_\nu 
  { d}x_{\mu_1}\cdots
  { d}x_{\mu_k}{ d}y_\nu{ d}y_{\nu_1}\cdots { d}y_{\nu_l}~.
\end{eqnarray}
Divergence operators $d^*_x$ (\ref{diver}) and $d^*_y$  may also be extended to $(k,l)$-forms:
\begin{eqnarray}
  \label{diverxy}
  &&{ d}_x^\ast f=\frac{1}{(k-1)!l!}\partial^\ast_\mu 
  f_{\mu\mu_2\cdots\mu_k;\nu_1\cdots\nu_l}(x,y)
  {d}x_{\mu_2}\cdots
  {d}x_{\mu_k}{ d}y_{\nu_1}\cdots { d}y_{\nu_l}~, \nonumber\\
  && { d}_y^\ast f=\frac{(-1)^k}{k!(l-1)!} 
  f_{\mu_1\cdots\mu_k;\nu\nu_2\cdots\nu_l}(x,y)\overleftarrow\partial{}^\ast_\nu 
  { d}x_{\mu_1}\cdots
  { d}x_{\mu_k}{ d}y_{\nu_2}\cdots {d}y_{\nu_l}~. 
\end{eqnarray}
It is straightforward to show that these operators satisfy nilpotency  
and anticommutation relations between them, when they act on the ``bi-local" form space $\Omega_{k,l}$,
\begin{eqnarray}
  \label{dxdy}
  &&{ d}_x^2={ d}_y^2={ d}_x^{\ast2}={ d}_y^{\ast2}=0~,\nonumber\\
  ({ d}_x+{ d}_y)^2&=&({ d^*}_x+{ d}_y)^2=({ d}_x+{ d^*}_y)^2=({ d}^\ast_x+{ d}^\ast_y)^2=0~.
\end{eqnarray}
There are four pairs of anticommuting nilpotent operators as $\{d_x, d_y\}$,$\{d_x, d^*_y\}$,
$\{d^*_x, d_y\}$ and $\{d^*_x, d^*_y\}$, any one of them can be use to construct a double complex. 

For  example, an $(k,l)$ differential form $\omega^{k,l}(x,y)$ satisfying 
\begin{eqnarray}
  \label{dxdyw}
  {d}_x^\ast{d}_y^\ast\omega^{k,l}(x,y)=0
\end{eqnarray}
then by the Poincar\'e 
lemma there exist form $\omega^{k-1,l+1}(x,y)$ satisfying 
\begin{eqnarray}
  \label{eq:dxwdyw}
    {d}_x^\ast\omega^{k,l}(x,y)+{d}_y^\ast\omega^{k-1,l+1}(x,y)=0~.
\end{eqnarray}
Since form $\omega^{k-1,l+1}(x,y)$ also satisfy (\ref{dxdyw}), they lead to new form 
$\omega^{k-2,l+2}(x,y)$. Such procedure can be continued until one ends up with 
$\omega^{0,k+l}(x,y)$ for $k+l\leq D$. In this way we got a double complex for $\{d^*_x, d^*_y\}$.
These formulas are also called descent equations, which 
were used  \cite{FSW1} in analyzing the
chiral anomaly in arbitrary even dimensional lattice gauge theory. 
In this paper we want to use another double complex in terms of 
operators $\{d_x, d^*_y\}$ to discuss the chiral anomaly.

\section{Double complex, descent equation and its solutions}
\setcounter{equation}{0}

The technique of double complex and its descent equation are widely used in the anomaly 
analysis of continuous gauge field theory.
It is our purpose to extend this technique to the anomaly analysis 
in lattice gauge theory using the NCDG on the lattice. As we did in \cite {FSW1} in terms of 
$\{d^*_x, d^*_y\}$ double complex, here we will use  $\{d_x, d^*_y\}$ double complex to do 
anomaly analysis and to derive the chiral anomaly of abelian lattice gauge theory 
 in arbitrary even dimensions.

We focus now on the ``bi-local" fields space $\bigcup_{k,l=0,1,\cdots,D}\Omega_{k,l}$ 
defined on $\rm Z^D\times\rm Z^D$ and 
start from an $(D,2)$-form $\omega^{D,2}(x,y)$ which satisfies 
\begin{eqnarray}
  \label{ desc  }
    \sum_x d^*_y \omega^{D,2}(x,y)=0.
\end{eqnarray}
Since $d^*_y \omega^{D,2}(x,y)$ is an $D$-form for $x$ in terms of Poincar\'e lemma there exist 
a $(D-1,1)$-form $\omega^{D-1,1}(x,y)$ such that 
\begin{eqnarray}
  \label{desc1}
    d^*_y \omega^{D,2}(x,y) + d_x\omega^{D-1,1}(x,y) = 0.
\end{eqnarray}
Acting $d^*_y$ on the above equation we get 
~   $ d^*_y d_x\omega^{D-1,1}(x,y) = 0,$ which is equivalent to $ d_xd^*_y \omega^{D-1,1}(x,y) = 0$. 
Then using  Poincar\'e lemma for $d_x$ again, there exist an $(D-2,0)$-form $\omega^{D-2,0}(x,y)$ 
such that
\begin{eqnarray}
  \label{desc2}
    d^*_y \omega^{D-1,1}(x,y) + d_x\omega^{D-2,0}(x,y) = 0.
\end{eqnarray}
These are the descent equations for $\omega^{D,2}(x,y)$. For general $\omega^{D,l}(x,y) $
which satisfies 
\begin{eqnarray}
  \label{desc3}
    \sum_x d^*_y \omega^{D,l}(x,y)=0.
\end{eqnarray}
one could get the following set of descent equations:
\begin{eqnarray}
  \label{desc4}
        d^*_y \omega^{D,l}(x,y) &+& d_x\omega^{D-1,l-1}(x,y) = 0, \nonumber\\
        d^*_y \omega^{D-1,l-1}(x,y)& +& d_x\omega^{D-2,l-2}(x,y) = 0, \nonumber\\
&               &\cdots,\nonumber\\
        d^*_y \omega^{D-l+1,1}(x,y)& +& d_x\omega^{D-l,0}(x,y) = 0.
\end{eqnarray}
It should be mentioned, according to the Poicar\'e Lemma\cite{Lus2, FSW1}, that if $\omega^{D,l}(x,y)$ is gauge invariant and locally depending on gauge field 
$A_\mu$ then all $\omega^{D-m,l-m}(x,y) ~(m=1,2,\cdots,l)~$ appeared in (\ref{desc4}) 
are gauge invariant and locally 
depending on $A_\mu$.

Now we are going to solve the descent equation. We start from $\omega^{D,l}(x,y)$  which is a $D$-form
in $x$ and $l$-form in $y$, its dual version is $0$-form in $x$ and $(D-l)$-form in $y$. 
We can use 
the dual version expression for the simplicity to express the $\omega^{D,l}(x,y)$ as,
\begin{eqnarray}
        \label{omega}
        \omega^{D,l}(x,y)=\frac {1} {l!}dx_{\mu_1}\cdots dx_{\mu_l}
dy_{\mu_1}\cdots dy_{\mu_l}\omega_{\mu_{l+1},\cdots,\mu_{D}}(x,y)dx_{\mu_{l+1}}\cdots dx_{\mu_D}.
\end{eqnarray}
The position of coefficient in the right side of (\ref{omega}) is of specific meaning. According to 
$ {\bf {Lemma 1}}$ one move the position of coefficients cross the forms, 
the value of argument in coefficients will shift according to the noncommutavity of forms and functions 
(\ref{ncomm}). So if two similar expression of forms have different order in their coefficient and form basis, 
then they absolutely different from each other.
 The position of coefficient
$\omega_{\mu_{l+1},\cdots,\mu_{D}}(x,y)$ in right side of (\ref{omega}) is chosen such that it will be  
more convenient for us in the following discussion to reach our final result.

In order to get the $\omega^{D-1,l-1}(x,y)$, firstly we introduce $\alpha^{D,l}(y)$ as, 
\begin{eqnarray}
        \label{alpha}
        \alpha^{D,l}(y)&=&\sum_x \omega^{D,l}(x,y) \nonumber\\
&=&\frac {1} {l!}dx_{\mu_1}\cdots dx_{\mu_l}
dy_{\mu_1}\cdots dy_{\mu_l}\alpha_{\mu_{l+1},\cdots,\mu_{D}}(y)dx_{\mu_{l+1}}\cdots dx_{\mu_D}.
\end{eqnarray}
where $\alpha_{\mu_{l+1},\cdots,\mu_{D}}(y) = \sum _x\omega_{\mu_{l+1},\cdots,\mu_{D}}(x,y)$.
 $\alpha^{D,l}(y)$ is the most $d_x$ cohomological non-trivial part .
Since the  $\sum_x[\omega^{D,l}(x,y) - \delta_{x,y}\alpha^{D,l}(x)]=0$  for a $D$-form for $x$,
by Poicar\'e Lemma for $d_x$ we get,
\begin{eqnarray}
        \label{omegadl}
        \omega^{D,l}(x,y)=\delta_{x,y}\alpha^{D,l}(x)+d_x\vartheta^{D-1,l}(x,y).
\end{eqnarray}
Substituting this result  into the first formula in (\ref{desc4}) one obtains, 
\begin{eqnarray}
        \label{omegadl1}
        d_x \omega^{D-1,l-1}(x,y)= - d^*_y\delta_{x,y}\alpha^{D,l}(x)+d_xd^*_y\vartheta^{D-1,l}(x,y).
\end{eqnarray}
The final expression for $\omega^{D-1,l-1}(x,y)$ will be given soon after the following lemma,

\vskip .5cm
\noindent{\bf Lemma 5:} If the $\alpha^{D,l}(x)$ is given in (\ref{alpha}), 
there is a $(D-1,l-1)$-form $\alpha^{D-1,l-1}(x)$ satisfying,
\begin{eqnarray}
  \label{dydx}
-d^*_y[\delta_{x,y}\alpha^{D,l}(x)]=d_x[\delta_{x,y}\alpha^{D-1,l-1}(x)],
\end{eqnarray}
where $\alpha^{D-1,l-1}(x)$ is given by
\begin{eqnarray}
        \label{alpha1 }
        \alpha^{D-1,l-1}(x)
&=&\frac {1} {(l-1)!}dx_{\mu_2}\cdots dx_{\mu_l}
dy_{\mu_2}\cdots dy_{\mu_l}\alpha_{\mu_{l+1},\cdots,\mu_{D}}(x)dx_{\mu_{l+1}}\cdots dx_{\mu_D}.
\end{eqnarray}
\vskip .5cm
\noindent{\bf Proof:} As $\sum_x d_x f(x,y)$ vanishes for any local field $f(x,y)$,
taking summation over $x$ of the first equation in (\ref{desc4}) one can get 
$\sum_x d^*_y \omega^{D,l}(x,y) =0$ .
Changing the order of $\sum_x$ and $d^*_y$ we get an important property of $\alpha^{D,l}(y) $, 
\begin{eqnarray}
        \label{dystar}
        d^*_y \alpha^{D,l}(y)&=&0.
\end{eqnarray}
This means that
\begin{eqnarray}
        \label{dystar1}
\frac {1} {l!}dx_{\mu_1}\cdots dx_{\mu_l}
dy_{\mu_2}\cdots dy_{\mu_l}[\alpha_{\mu_{l+1},\cdots,\mu_{D}}(y+\widehat{\mu_1}    )-
\alpha_{\mu_{l+1},\cdots,\mu_{D}}(y)]
dx_{\mu_{l+1}}\cdots dx_{\mu_D}=0.
\end{eqnarray}
Then expand the both sides of (\ref{dydx}) by definition of $d_x$ and $d^*_y$,
\begin{eqnarray}
        \label{}
d^*_y[\delta_{x,y}\alpha^{D,l}(x)]&=&\frac {1} {(l-1)!}[\delta_{x,y}dx_{\mu_1}\cdots dx_{\mu_l}
dy_{\mu_2}\cdots dy_{\mu_l}\alpha_{\mu_{l+1},\cdots,\mu_{D}}(x)dx_{\mu_{l+1}}\cdots dx_{\mu_D}-\nonumber\\
&&\delta_{x,y-\widehat{\mu_1}}dx_{\mu_1}\cdots dx_{\mu_l}
dy_{\mu_2}\cdots dy_{\mu_l}\alpha_{\mu_{l+1},\cdots,\mu_{D}}(x)dx_{\mu_{l+1}}\cdots dx_{\mu_D}] \nonumber\\
d_x[\delta_{x,y}\alpha^{D-1,l-1}(x)]&=&\frac {1} {(l-1)!}[\delta_{x+\widehat{\mu_1},y}
dx_{\mu_1}\cdots dx_{\mu_l}
dy_{\mu_2}\cdots dy_{\mu_l}\alpha_{\mu_{l+1},\cdots,\mu_{D}}(x)dx_{\mu_{l+1}}\cdots dx_{\mu_D}-\nonumber\\
&&\delta_{x,y}dx_{\mu_1}\cdots dx_{\mu_l}
dy_{\mu_2}\cdots dy_{\mu_l}\alpha_{\mu_{l+1},\cdots,\mu_{D}}(x-\widehat{\mu_1})dx_{\mu_{l+1}}
\cdots dx_{\mu_D}] \nonumber\\
\end{eqnarray}
Using (\ref{dystar1}) and the identity $\delta_{x+\widehat{\mu_1}, y}=\delta_{x, y-\widehat{\mu_1}}$, 
one can easily prove this lemma. 

\vskip .5cm

From this lemma and (\ref{omegadl1}), it is easy to find,
\begin{eqnarray}
        \label{omega2}
        d_x[ \omega^{D-1,l-1}(x,y) - \delta_{x,y}\alpha^{D-1,l-1}(x)-d^*_y\vartheta^{D-1,l}(x,y)]=0.
\end{eqnarray}
Then we obtain the expression for $\omega^{D-1,l-1}(x,y) $ 
\begin{eqnarray}
        \label{omega3}
        \omega^{D-1,l-1}(x,y) = \delta_{x,y}\alpha^{D-1,l-1}(x)+d^*_y\vartheta^{D-1,l}(x,y)+
d_x\vartheta^{D-2,l-1}(x,y).
\end{eqnarray}

In the same way,  we can find,
\begin{eqnarray}
        \label{omega4}
        \omega^{D-2,l-2}(x,y) = \delta_{x,y}\alpha^{D-2,l-2}(x)+d^*_y\vartheta^{D-2,l-1}(x,y)+
d_x\vartheta^{D-3,l-2}(x,y),
\end{eqnarray}
where
\begin{eqnarray}
        \label{alpha2 }
        \alpha^{D-2,l-2}(x)
&=&\frac {1} {(l-2)!}dx_{\mu_3}\cdots dx_{\mu_l}
dy_{\mu_3}\cdots dy_{\mu_l}\alpha_{\mu_{l+1},\cdots,\mu_{D}}(x)dx_{\mu_{l+1}}\cdots dx_{\mu_D}.
\end{eqnarray}
The general solution for descent equation will be obtained similarly,
\begin{eqnarray}
        \label{omega5}
        \omega^{D-k,l-k}(x,y) = \delta_{x,y}\alpha^{D-k,l-k}(x)+d^*_y\vartheta^{D-k,l-k+1}(x,y)+
d_x\vartheta^{D-k-1,l-k}(x,y),
\end{eqnarray}
where
\begin{eqnarray}
        \label{alpham }
        \alpha^{D-k,l-k}(x)
&=&\frac {1} {(l-k)!}dx_{\mu_{k+1}}\cdots dx_{\mu_l}
dy_{\mu_{k+1}}\cdots dy_{\mu_l}\alpha_{\mu_{l+1},\cdots,\mu_{D}}(x)dx_{\mu_{l+1}}\cdots dx_{\mu_D},
\end{eqnarray}
for $k=1,2.\cdots, l$.

\section{Axial anomaly of abelian lattice gauge theory}
\setcounter{equation}{0}

As mentioned in Section 1, the chiral anomaly (\ref{anom}) in abelian lattice gauge theory, 
is gauge invariant, locally dependent on gauge field $A_\mu$ and its variation with respect to 
$A_\mu$ satisfies (\ref{topo})
which reflects the topological nature of the anomaly. Then its topological part will be fixed 
up to some constant coefficient. The result  in the case of four-dimensions is given by 
L\"uscher's theorem (\ref {anom1}), 
its extension to higher-dimensions could be found in \cite {FSW, FSW1}. Now we use the new double complex 
and descent equation 
proposed in the last section to derive it again. First we state the main result.

\vskip .5cm
\noindent{\bf Theorem 1:} In $D$-dimensional infinite hypercubic lattice, if the anomaly 
written in $D$-form $Q(x)=q(x)d^Dx$ is gauge invariant, locally depending on $U(1)$ gauge field $A_\mu$ 
and of topological nature 
\begin{eqnarray}
  \label{natu}
\sum_x \delta Q(x)=0 
\end{eqnarray}
 then, 
\begin{eqnarray}
  \label{eq:anom}
    Q(x)&=& \sum_{n=0}^{[D/2]}F^n B^{(D-2n)} +dK^{(D-1)}
\end{eqnarray}
where $F$ is the gauge field strength $2$-form, $B^{(D-2n)}$ is constant $(D-2n)$-form and 
$K^{(D-1)}$ 
is gauge invariant $(D-1)$-form locally depending on $A_\mu$.
\vskip .5cm

One can easily find the theorem is consistent with result in \cite{FSW, FSW1}, 
\begin{eqnarray}
q(x)&=\alpha
   +\sum_{n=1}^{[D/2]}
   \beta_{\mu_1\nu_1\mu_2\nu_2\cdots\mu_n\nu_n}
   F_{\mu_1\nu_1}(x)
   F_{\mu_2\nu_2}(x+\widehat\mu_1+\widehat\nu_1)\cdots\times \nonumber\\
   & F_{\mu_n\nu_n}(x+\widehat\mu_1+\widehat\nu_1+\cdots
    +\widehat\mu_{n-1}+\widehat\nu_{n-1})+
   \partial_\mu^*k_\mu(x).
\end{eqnarray}
It is obvious one can recover L\"uscher's theorem (\ref {anom1}) exactly  when $D=4$.

The main theorem can be shown easily, once the following lemmas are obtained.
\vskip .5cm
\noindent{\bf Lemma 6:} On $D$-dimensional infinite hypercubic lattice, if 
$D$-form $Q(x)=q(x)d^Dx$ is gauge invariant, locally depending on $U(1)$ gauge field $A_\mu$ 
and of topological nature $\sum_x \delta Q(x)=0 $, then 
\begin{eqnarray}
   \label{eq:1}
        Q(x)&=&bd^Dx +F\alpha^{D-2}(x)+d_x\vartheta^{D-1}(x),
\end{eqnarray}
where $b$ is constant, $\alpha^{D-2}(x)$ and $\vartheta^{D-1}(x)$ are gauge invariant, 
locally depending on $A_\mu$, and $d \alpha^{D-2}(x) =0$.

\vskip .7cm
\noindent{\bf Lemma 7:} On $D$-dimensional infinite hypercubic lattice, if 
$2m$-form  $\alpha^{2m}(x)$ ($2m<D$) is gauge invariant, 
locally depending on $A_\mu$, and $d \alpha^{2m}(x) =0$, then 
\begin{eqnarray}
   \label{eq:2}
        \alpha^{2m}(x) &=&B^{2m} +F\alpha^{2m-2}(x)+d_x\vartheta^{2m-1}(x),
\end{eqnarray}
where $B^{2m}$ is constant $2m$ -form, $\alpha^{2m-2}(x)$ and $\vartheta^{2m-1}(x)$ are gauge invariant, 
locally depending on $A_\mu$, and $d \alpha^{2m-2}(x) =0$.
\vskip .5cm

If we accept these two lemmas, using the Bianchi identity $dF=0$ and the Leibniz rule 
for $d$ in the NCDG then substituting eq.(\ref{eq:2}) when $2m=D-2$ into eq. (\ref{eq:1}), 
one can easily find,
\begin{eqnarray}
        Q(x)&=&bd^Dx +FB^{D-2} +F^2\alpha^{D-4}(x)+d_x(F\vartheta^{D-3}(x)+\vartheta^{D-1}(x)),
\end{eqnarray}
Taking $2m=D-4$ in {\bf Lemma 7} 
substituting $\alpha^{D-4}(x)$ in the above formula again and repeating this 
procedure up to $2m=2$, we will arrive at the main {\bf Theorem}.

Now we try to prove lemma 6 and 7.
\vskip .5cm
\noindent{\bf Proof of Lemma 6:} Following the discussion in \cite{Lus2, FSW1} 
some algebraic techniques could be 
used to separate $Q(x)$ into two parts, one is independent of gauge field $A_\mu$ and other dependent 
on  gauge field $A_\mu$. Also we can assume the first part is a constant, as the translation 
invariance of lattice gauge 
theory force all the dependence on $x$ should be appeared through gauge field $A_\mu$.
 \begin{eqnarray}
        \label{qx}
Q(x)
&=&bd^Dx +\int_0^1 dt \Biggl(\frac{\partial Q(x)}{\partial t}\Biggr)_{A\rightarrow tA}\nonumber\\
&=&bd^Dx +\sum_y \int_0^1 dt \Biggl(\frac{\delta Q(x)}{\delta A_\nu(y)}\Biggr)_{A\rightarrow tA}A_\nu(y)\nonumber\\
&=&bd^Dx +\sum_y <J^{D,1}(x,y), A(y)>,
\end{eqnarray}
where $J^{D,1}(x,y)$ is an $(D,1)$-form in $\rm Z^D\times\rm Z^D$ as,
\begin{eqnarray}
J^{D,1}(x,y)=\int_0^1 dt \Biggl(\frac{\delta Q(x)}{\delta A_\nu(y)}\Biggr)_{A\rightarrow tA}dy_\nu,
\end{eqnarray}
 and 
$A(y)$ is gauge field $1$-form.

As in the assumption $Q(x)$ locally depends on the gauge field $A_\mu(x)$, 
the meaning of locality is nothing but the  $J^{D,1}(x,y)$ decrease 
at least exponentially as $|x-y|\to\infty$. Then the summation over $y$ in (\ref{qx}) is finite.
Since the gauge group is abelian, the variation of gauge invariant 
field with respect to gauge potential is gauge invariant, as well as the ``bi-local" $J^{D,1}(x,y)$
 is gauge invariant.
Furthermore, we make an infinitesimal gauge transformation of $Q(x)$,
\begin{eqnarray}
 \delta_GQ(x) = 
\sum_y <J^{D,1}(x,y), d_y\lambda(y)>,
\end{eqnarray}
from the gauge invariance of  $Q(x)$, one get 
\begin{eqnarray}
d^*_y J^{D,1}(x,y)&=&0.
\end{eqnarray}
This means  
\begin{eqnarray}
J^{D,1}(x,y)&=&d^*_y\omega^{D,2}(x,y),
\end{eqnarray}
where $\omega^{D,2}(x,y)$ is a ``bi-local" gauge invariant $(D,2)$-form. Then we have,
 \begin{eqnarray}
        \label{qx1}
Q(x)
&=&bd^Dx +\sum_y <\omega^{D,2}(x,y), F(y)>,
\end{eqnarray}
where the dual property of $-d_y$ and $d^*_y$ are used. The topological nature (\ref{topo}) of $Q(x)$ tell us,
\begin{eqnarray}
\sum_x J^{D,1}(x,y)&=&\sum_x\int_0^1 dt \Biggl(\frac{\delta Q(x)}{\delta A_\nu(y)}\Biggr)_{A\rightarrow tA}
dy_\nu \nonumber\\
    &=&\int_0^1 dt \sum_x \Biggl(\frac{\delta Q(x)}{\delta A_\nu(y)}\Biggr)_{A\rightarrow tA}dy_\nu \nonumber\\
&=&0.
\end{eqnarray}
That shows us that  $\omega^{D,2}(x,y)$ satisfies the descent equation,
\begin{eqnarray}
\label{ome}
\sum_x d^*_y\omega^{D,2}(x,y)&=&0.
\end{eqnarray}
using the general solution to descent equation (\ref{omegadl}), taking $l=2$, i.e.
\begin{eqnarray}
        \omega^{D,2}(x,y)&=&\delta_{x,y}\alpha^{D,2}(x)+d_x\vartheta^{D-1,2}(x,y),\nonumber\\
\alpha^{D,2}(x)
&=&\frac {1} {2!}dx_{\mu_1}dx_{\mu_2}
dy_{\mu_1} dy_{\mu_2}\alpha_{\mu_{3},\cdots,\mu_{D}}(x)dx_{\mu_{3}}\cdots dx_{\mu_{D}},
\end{eqnarray}
and substituting them into (\ref{qx1}), it will prove the lemma,
\begin{eqnarray}
        \label{}
Q(x)
&=&bd^Dx +\sum_y <\delta_{x,y}\alpha^{D,2}(x)+d_x\vartheta^{D-1,2}(x,y), F(y)>,\nonumber\\
&=&bd^Dx +\sum_y <\delta_{x,y}\frac {1} {2!}dx_{\mu_1}dx_{\mu_2}
dy_{\mu_1} dy_{\mu_2}\alpha_{\mu_{3},\cdots,\mu_{D}}(x)dx_{\mu_{3}}\cdots dx_{\mu_{D}},
 F_{\rho_1\rho_2}dy_{\rho_1}dy_{\rho_2}>\nonumber\\
&&+d_x\sum_y <\vartheta^{D-1,2}(x,y), F(y)>\nonumber\\
&=&bd^Dx +F\alpha^{D-2}(x)+d_x\vartheta^{D-1}(x),
\end{eqnarray}
where $(D-2)$-form $\alpha^{D-2}(x)= \frac {1} {2!}\alpha_{\mu_{3},\cdots,\mu_{D}}(x)dx_{\mu_{3}}\cdots 
dx_{\mu_{D}}$ and 
 $(D-1)$-form $\vartheta^{D-1}(x)=\sum_y <\vartheta^{D-1,2}(x,y), F(y)>$ are gauge invariant local field. 
From the Leibniz rule in NCDG one can easily prove $(D-2)$-form $\alpha^{D-2}(x)$ is closed.

\vskip .3cm
\noindent{\bf Proof of Lemma 7:} The proof is much similar to that of {\bf  Lemma 6}. First we 
 separate $\alpha^{2m}(x)$ into two parts, one is independent of gauge field $A_\mu$ and the other depends 
on  gauge field $A_\mu$,
 \begin{eqnarray}
        \label{alpha2m}
\alpha^{2m}(x)
&=&B^{2m} +\int_0^1 dt \Biggl(\frac{\partial \alpha^{2m}(x)}
{\partial t}\Biggr)_{A\rightarrow tA}\nonumber\\
&=&B^{2m} +\sum_y \int_0^1 dt \Biggl(\frac{\delta \alpha^{2m}(x)
}{\delta A_\nu(y)}\Biggr)_{A\rightarrow tA}A_\nu(y)\nonumber\\
&=&B^{2m} +\sum_y <J^{2m,1}(x,y), A(y)>,
\end{eqnarray}
where $J^{2m,1}(x,y)$ is an $(2m,1)$-form in $\rm Z^D\times\rm Z^D$ as,
\begin{eqnarray}
J^{2m,1}(x,y)=\int_0^1 dt \Biggl(\frac{\delta \alpha^{2m}(x)
}{\delta A_\nu(y)}\Biggr)_{A\rightarrow tA}dy_\nu.
\end{eqnarray}
 
From the locality of $\alpha^{2m}(x)
$, we know the $J^{2m,1}(x,y)$ is a ``bi-local" field i.e. it decreases 
at least exponentially as $|x-y|\to\infty$ and the summation over $y$ in (\ref{alpha2m}) is finite.
Since $\alpha^{2m}(x)$
is  gauge invariant 
field one can get, 
\begin{eqnarray}
d^*_y J^{2m,1}(x,y)&=&0.
\end{eqnarray}

From Paincar\'e lemma it follows, 
\begin{eqnarray}
J^{2m,1}(x,y)&=&d^*_y\omega^{2m,2}(x,y),
\end{eqnarray}
where $\omega^{2m,2}(x,y)$ is a gauge invariant $(2m,2)$-form ``bi-local" field. Then we have,
 \begin{eqnarray}
        \label{aalpha2m}
\alpha^{2m}(x)
&=&B^{2m} +\sum_y <\omega^{2m,2}(x,y), F(y)>.
\end{eqnarray}

 The closedness of $\alpha^{2m}(x)$
 tell us,
\begin{eqnarray}
d_xd^*_y\omega^{2m,2}(x,y)= d_x J^{2m,1}(x,y)=0.
\end{eqnarray}
That shows us that  $\omega^{2m,2}(x,y)$ satisfies the descent equation.
Using the general solution to descent equation (\ref{omega5}), 
\begin{eqnarray}
        \omega^{2m,2}(x,y)&=&\delta_{x,y}\alpha^{2m,2}(x)+d_x\vartheta^{2m-1,2}(x,y)+
d^*_y\vartheta^{2m,3}(x,y),\nonumber\\
\alpha^{2m,2}(x)
&=&\frac {1} {2!}dx_{\mu_1}dx_{\mu_2}
dy_{\mu_1} dy_{\mu_2}\alpha_{\mu_{3},\cdots,\mu_{2m}}(x)dx_{\mu_{3}}\cdots dx_{\mu_{2m}}.
\end{eqnarray}
Substituting them into (\ref{aalpha2m}), we prove the lemma,
\begin{eqnarray}
\alpha^{2m}(x)
&=&B^{2m} +\sum_y <\delta_{x,y}\alpha^{2m,2}(x)+d_x\vartheta^{2m-1,2}(x,y), F(y)>,\nonumber\\
&=&B^{2m} +\sum_y <\delta_{x,y}\frac {1} {2!}dx_{\mu_1}dx_{\mu_2}
dy_{\mu_1} dy_{\mu_2}\alpha_{\mu_{3},\cdots,\mu_{2m}}(x)dx_{\mu_{3}}\cdots dx_{\mu_{2m}},
 F_{\rho_1\rho_2}dy_{\rho_1}dy_{\rho_2}>\nonumber\\
&&+d_x\sum_y <\vartheta^{2m-1,2}(x,y), F(y)>\nonumber\\
&=&B^{2m} +F\alpha^{2m-2}(x)+d_x\vartheta^{2m-1}(x),
\end{eqnarray}
where $(2m-2)$-form $\alpha^{2m-2}(x)= \frac {1} {2!}\alpha_{\mu_{3},\cdots,\mu_{2m}}(x)
dx_{\mu_{3}}\cdots dx_{\mu_{2m}}$ and 
 $(2m-1)$-form $\vartheta^{2m-1}(x)=\sum_y <\vartheta^{2m-1,2}(x,y), F(y)>$ are gauge invariant local field. 
From the Leibniz rule for NCDG one can easily prove $(2m-2)$-form $\alpha^{2m-2}(x)$ is closed.

\section{Conclusion}
\setcounter{equation}{0}

We would like to make a few remarks in this conclusion. 

We have extended the axial anomaly of ref. \cite{Lus2} in four dimensions to arbitrary 
even dimensions in abelian lattice gauge theory. In the derivation of chiral anomaly
in this  paper as well as in \cite {Lus2, FSW, FSW1},  the locality plays a very important role. 
The locality here means,
if taking  the continuous limit one should get the correct continuous field theory, so 
we should keep this property 
in every step to make our derivation reliable.

We have given four different kinds of double complex in section 2. Two of them have been used to 
the anomaly analysis in abelian lattice gauge theory, one is in \cite{FSW1} and second in this paper. 
The application of the other two double complex is also very interesting, which is under considerations.

 Lattice gauge theory is a kind of discretized 
field theory. The discreteness in this theory results many specific characters 
of lattice gauge theory from the  usual continuum field theory. Such as the gauge field on 
lattice gauge theory is given by parallel transporter other than gauge vector field, 
the chirality as well as fermion doubling in lattice gauge theory are more serious problems
and trouble us many years even up to now. Discreteness could be represented by NCDG in lattice 
and very recently NCDG is used in lattice gauge theory to understand some quantum effect or topological 
effect such as chiral anomaly in \cite {Lus2, FSW, FSW1} and this paper. 
However it is not hard to notice that in our discussion, we use many geometric concepts and 
tools step from ordinary commutative differential geometry. For example, the concept of double 
complex and descent equation lead us to the the calculation of chiral anomaly in lattice gauge theory 
in the same spirit of anomaly analysis in continuum quantum field theory, although the exact 
meaning of them is different essentially. 
We hope that we can 
understand more about lattice theory along this way from point of view of NCDG in the near future. 

\vskip 1cm

One of us (K.W.) is very grateful to
F.~Sakata for the warm hospitality extended to him and to Faculty of
Science of Ibaraki University for the financial support during his
visiting at Ibaraki University. He would like to express his thanks also to the organizers of the workshop 
``Noncommutative Geometry and its Application to Physics" 
hold at Shonan-kokusaimura on May 31st to June 4th of 1999
for inviting him to join this workshop.


\end{document}